\documentstyle[aps,prd,eqsecnum,epsfig,subfigure,amssymbols]{revtex}
\begin{document}
\draft

\thispagestyle{empty}
{\baselineskip0pt
\rightline{\large\baselineskip16pt\rm\vbox to20pt{\hbox{YITP-01-22}
               \hbox{WU-AP/121/01}
\vss}}%
}

\begin{center}
{\large {Physical aspects of naked singularity explosion\\
{\it -- How does a naked singularity explode? --}} \\~\\

{Hideo Iguchi\dag\footnote{Present address: Department of Physics, Tokyo Institute of Technology Oh-Okayama, Meguro-ku, Tokyo 152-8550, Japan} ~ and  Tomohiro Harada\ddag} \\~\\}

{\dag\ Yukawa Institute for Theoretical Physics, Kyoto University, Kyoto
606-8502, Japan} \\
{\ddag\ Department of Physics, Waseda University, Shinjuku, Tokyo
169-8555, Japan}  
%
\end{center}

\begin{abstract}                
The behaviors of quantum stress tensor for the scalar field on the
classical background of spherical dust collapse is studied. 
In the previous works diverging flux of quantum radiation was
predicted. We use the 
exact expressions in a 2D model formulated by Barve {\it et al}.
Our present results show that the back reaction does not become
important during the semiclassical phase. The appearance of the naked
singularity would not be affected by this quantum field radiation. To
predict whether the naked singularity explosion occurs or not we need
the theory of quantum gravity.  
We depict the generation of the diverging flux inside the collapsing star.
The quantum energy is gathered around the
center positively. This would be converted to the diverging flux along
the Cauchy horizon. The ingoing negative flux crosses the
Cauchy horizon. The intensity of it is divergent only at the central
naked singularity. This diverging negative ingoing flux is balanced with 
the outgoing positive diverging flux which 
propagates along the Cauchy horizon.
After the replacement of the naked singularity to the practical high
density region the instantaneous diverging radiation would change to 
more milder one with finite duration. 
\end{abstract}

\section{Introduction}
It is known that several models of 
gravitational collapse end in naked singularities from regular initial
data. Among these models the most fully studied one is the spherically
symmetric inhomogeneous dust collapse, which is described by the
Lema\^{\i}tre-Tolman-Bondi (LTB) solution 
\cite{Lemaitre:1933,Tolman:1934,Bondi:1947}. 
The final fate of this model was investigated by numerous authors
\cite{Eardley:1979tr,Christodoulou:1984,Newman:1986,Joshi:1993zg,Jhingan:1997ia}.
They concluded that the singularity could be naked or covered by event 
horizon depending on the initial density and velocity profiles of
collapsing cloud.
There are also other examples of the naked singularity
formation. Recent review of these examples and cosmic censorship
hypothesis were given by some authors \cite{Wald:1997wa,Penrose:1998}.

It is expected that the quantum effects analogous to Hawking radiation 
play an important role in the final stage of the naked singularity formation. 
From this point of view some researchers investigated a quantized
massless scalar field on the classical background of such collapse
\cite{Ford:1978ip,Hiscock:1982pa,Barve:1998ad,Harada:2000jy,Harada:2000ar,Singh:2000sp}.
Their results showed that the diverging outgoing quantum
radiation appears in the approach to the Cauchy horizon.
These results are in contrast to the investigation of the  
classical radiation, i.e., gravitational waves given by the authors and 
K. Nakao\cite{Iguchi:1998qn,Iguchi:1999ud,Iguchi:2000jn,Nakao:2000as}, 
where the energy flux of the
gravitational waves remains finite as the Cauchy horizon is approached. 
Therefore the interpretation of this diverging quantum flux is 
important for the understanding of the nature of these naked
singularities and for the proof or disproof of the cosmic censorship
hypothesis. 
In another context we can also expect the intense
radiation from a naked singularity and get information about 
strongly curved spacetime. In this respect the naked singularity
formation is discussed as a possible origin of $\gamma$-ray bursts
\cite{Harada:2000ar,Joshi:2000ht}. Also these observational aspects of
the naked singularity might be useful to distinguish naked
singularities from black holes on the astrophysical observations
\cite{Vaz:1998gd,Virbhadra:1998}.

The above quantum radiation was estimated perturbatively in the
semiclassical treatment. Seemingly the divergence of the quantum flux
would mean the breakdown of the perturbative estimate and the
plausible next step will be to consider the backreaction of the
quantum radiation. However this is not necessarily true. At the
Cauchy horizon the classical gravity is broken down because of the
causal connection to the singularity. Therefore the
semiclassical treatment, where the gravity is treated as classical, is
also broken down independently of
whether or not the quantum radiation diverges.
It is not obvious whether the backreaction of the quantum
radiation is significant or not within the region where the semiclassical
treatment is plausible. 

There would be two criterion
for deciding when the back reaction becomes important. One is
that the total energy received at infinity becomes comparable to the
mass of the collapsing star. In \cite{Harada:2000me} it has been shown 
that, in the spherical dust model, the emitted flux is at most 
order of Planck energy before the quantum gravity phase. 
Here the quantum gravity phase means the spacetime region to where the
central high curvature region can causally affect, e.g., the region
within one Planck time up to the Cauchy horizon. If the mass
of the collapsing star is much greater than the Planck mass, then the
back reaction does not become important. The second criterion is that
the energy density of the quantum field becomes comparable to the
background energy density inside the star. To estimate the energy
density of the quantum field we should calculate the vacuum
expectation value of the stress tensor for the quantum field. In a
four-dimensional model, it is difficult to establish this
issue. However one can study the quantum stress tensor for a
two-dimensional model which is obtained by suppressing the angular
part of the 4D model \cite{Davies:1976ei}. 
For the self-similar dust collapse, the exact expression of it
was computed \cite{Barve:1998tv}. Using this model, this paper
investigates the exact behaviors of the quantum stress tensor and
discusses the generation of the diverging quantum flux inside the star.
Also we show that the back reaction does not become significant during the
semiclassical evolution inside the collapsing 
star fated to be a naked singularity.

The plan of this paper is as follows. In Sec. \ref{sec:Exact expression} 
we introduce the calculation for the quantum stress tensor for a
massless scalar field in the 2D self-similar spherical dust collapse
performed by Barve {\it et al.} \cite{Barve:1998tv}.
In Sec. \ref{sec:Exact behavior} using this expressions we investigate 
the exact behaviors of the quantum stress tensor and consider 
the generation of diverging flux and the  criterion of the 
semiclassical treatment inside the cloud.
In Sec. \ref{sec:Summary} we
summarize our results. We use the units $G=c=\hbar=1$.

\section{Exact expression for quantum stress tensor in self-similar
dust collapse}
\label{sec:Exact expression}

We consider 
the exact expression for the quantum stress tensor of the massless
scalar fields in a 2D collapsing LTB spacetime. This was first done by Barve 
{\it et al.} \cite{Barve:1998tv}.  Here we briefly introduce their
results, taking care of the consistency of signs. 
\subsection{Self-similar spherical dust collapse}
A spherically symmetric dust collapse is represented by the LTB solution.
Using the synchronous comoving coordinate system, the 
line element of the LTB spacetime can be expressed in the form  
\begin{equation}
 \label{metric}
 ds^2 = dt^2 - \tilde{A}^2(t,r) dr^2 - R^2(t,r) d\Omega^2.
\end{equation}
The energy-momentum tensor for the dust fluid is 
\begin{equation}
  \label{bgmatter}
  {T}^{\mu\nu} = {\rho}(t,r){u}^{\mu}{u}^{\nu}
\end{equation}
where ${\rho}(t,r)$ is the rest mass density and 
${u}^{\mu}$ is the 4-velocity of the dust fluid. 

Then the Einstein equations and the equation of motion for the 
dust fluid reduce to the following simple equations 
\begin{eqnarray}
  \tilde{A} &=&  \frac{R'}{\sqrt{1+f(r)}} \label{eq:A} \\
  {\rho}(t,r) &=& \frac{1}{8\pi}
  \frac{1}{R^2 R'}{dF(r)\over dr}\label{eq:einstein} \\
  \dot{R}^2-\frac{F(r)}{R} &=& f(r)\label{eq:energyeq} 
\end{eqnarray}
where $f(r)$ and $F(r)$ are arbitrary functions of the radial 
coordinate $r$, and the overdot and prime denote partial derivatives
with respect to $t$ and $r$, respectively. From equation (\ref{eq:einstein}), 
$F(r)$ is related 
to the Misner-Sharp mass function, $m(r)$, of the dust cloud in the manner 
\begin{equation}
  \label{mass}
  m(r) = 4\pi \int_0^{R(t,r)}{\rho}(t,r)R^2dR = 4\pi
  \int_0^r{\rho}(t,r)R^2R' dr =\frac{F(r)}{2}. 
\end{equation}
Hence equation (\ref{eq:energyeq}) might be 
regarded as the energy equation per unit mass. 
This means that  
the other arbitrary function, $f(r)$, is recognized as 
the specific energy of the dust fluid. 
The motion of the dust cloud is completely specified 
by the function, $F(r)$, and the specific energy, $f(r)$. 

The self-similar solution corresponds to the choice of 
$2m(r)=\lambda r$, where $\lambda$ is a constant, and $f(r)=0$.
Following Barve {\it et al.} \cite{Barve:1998tv}, we assume 
that the central singularity appears at 
$t=0$ and rescale $R$ as $R(0,r)$ $=$ $(3/2)^{2/3}\lambda^{1/3}r$ at $t=0$.
The self-similar collapse solution is then
\begin{equation}
 \label{R}
 R=(2m(r))^{1/3}\left(\frac{3}{2}(r-t)\right)^{2/3}.
\end{equation}
The derivative of $R$ with respect to $r$ is 
\begin{equation}
 \label{R'}
 R'=\left(\frac{9\lambda}{4}\right)^{1/3}\frac{1-z/3}{\left(1-z\right)^{1/3}}
\end{equation}
where $z=t/r$. And the first and second derivatives
of $R'$ with respect to $z$ are given by
\begin{equation}
 \label{dR'/dz}
 \frac{dR'}{dz}=\frac{2}{9}\left(\frac{9\lambda}{4}\right)^{1/3}
                \frac{z}{\left(1-z\right)^{4/3}}
\end{equation}
and
\begin{equation}
 \label{d^2R'/dz^2}
 \frac{d^2R'}{dz^2}=\frac{2}{9}\left(\frac{9\lambda}{4}\right)^{1/3}
                \frac{1+z/3}{\left(1-z\right)^{7/3}}.
\end{equation}
Using this solution, the density function $\rho(t,r)$ is expressed as
\begin{equation}
 \rho(t,r) = \frac{1}{6 \pi (r-t)(3r-t)}.
\end{equation}
At the center the density diverges as 
\begin{equation}
 \rho(t,0) = \frac{1}{6\pi t^2}
\end{equation}
to the singularity.

  From the analysis of the null geodesic equations, 
the condition for the central singularity to be naked is obtained as \cite{Joshi:1995br}
\begin{equation}
 \lambda \le  156 - 90\sqrt{3}=0.115427\cdots.
\end{equation}
If the singularity is naked, there exist $z$ which  satisfies
\begin{equation}
 z-R'(z)=z-\left(\frac{9\lambda}{4}\right)^{1/3}\frac{1-z/3}{(1-z)^{1/3}}=0.
\end{equation}
This condition is equivalent to the existence of positive root for the 
following equation,
\begin{equation}
 f(y) \equiv y^4 + \frac{a}{3} y^3 - y +\frac{2}{3}a =0 
\end{equation}
where $y=(1-z)^{1/3}$ and  $a=\left(\frac{9}{4}\lambda\right)^{1/3}$.
The Cauchy horizon is determined by $z=z_{\mbox{\tiny CH}}$ 
which satisfies this equation.
It can be shown that this naked singularity is always
globally naked \cite{Barve:1998ad}.

To obtain the exact expression for the quantum stress tensor, we need
double-null coordinates for the 2D part of the metric (\ref{metric}). 
In the interior region, we introduce null coordinates $(\eta, \zeta)$
\begin{equation}
 \eta=\left\{\begin{array}{ll} r e^{\int dz/(z-R')} & (z-R' > 0)\\
                         - r e^{\int dz/(z-R')} & (z-R' < 0)
           \end{array}\right. 
\end{equation}
and
\begin{equation}
 \zeta=\left\{\begin{array}{ll} r e^{\int dz/(z+R')} & (z+R' > 0)\\
                         - r e^{\int dz/(z+R')} & (z+R' < 0).
           \end{array}\right. 
\end{equation}
For the convenience of the calculation for the quantum stress tensor, 
we introduce another null coordinates $(U,V)$ as
\begin{eqnarray}
 U &=&\left\{\begin{array}{ll}
       \ln \eta &  (z-R' > 0)\\
        -\ln |\eta| & (z-R' < 0)\\
       \end{array}\right. \\
 V &=& \left\{\begin{array}{ll}
           \ln \zeta &  (z+R' > 0)\\
           - \ln |\zeta| & (z+R' < 0).
           \end{array}\right. 
\end{eqnarray}
In these coordinates the center of the cloud is expressed by $V=U$
 from the fact that $\eta=\zeta$ at the center \cite{Barve:1998ad}.
Hereafter we concentrate our attention to
the outside of the Cauchy horizon where $z-R'<0$. 
Then 2D metric is expressed as
\begin{equation}
\label{UVmetric}
 ds^2 = A^2(U,V) dU dV
\end{equation}
where 
\begin{equation}
 A^2(U,V) = \left\{\begin{array}{ll}
             -r^2 ( z^2 -R'^2) &  (z+R' > 0)\\
              r^2 ( z^2 -R'^2) &  (z+R' < 0).\\
           \end{array}\right.       
\end{equation}

In the exterior region, the metric can be expressed as 
\begin{equation}
 ds^2=D^2(u,v)dudv=\left(1-\frac{2M}{R}\right)dudv
\end{equation}
using  the Eddinton-Finkelstein double null coordinates given by
\begin{equation}
 \label{uv}
 u=T-R^*, \hspace{1cm} v= T+ R^*
\end{equation}
where
\begin{equation}
 \label{R*}
 R^* = R + 2 M \ln\left(\frac{R}{2M}-1\right)
\end{equation}
and
\begin{equation}
 \label{T}
 T = t -2 \sqrt{2MR} -2M \ln \frac{\sqrt{R}-\sqrt{2M}}{\sqrt{R}+\sqrt{2M}}.
\end{equation}

\subsection{Exact expression for quantum stress tensor}
For simplicity, we consider a minimally coupled scalar field
as a quantum field, although the situation would not be changed for 
other massless fields, such as electromagnetic fields.

It is known that any two dimensional spacetime is conformally flat. 
Then its metric can be expressed by double null coordinates 
$(\hat{u},\hat{v})$ as
\begin{equation}
 ds^2 = C^2(\hat{u},\hat{v})d\hat{u}d\hat{v}.
\end{equation}
If the initial quantum state is set to the vacuum state in the 
Minkowski spacetime $ds^{2}=d\hat{u}d\hat{v}$, then
the expectation value of the stress tensor of the scalar field is
given by \cite{Davies:1976ei}
\begin{eqnarray}
 \langle T_{\hat{u}\hat{u}}\rangle 
  &=& -\frac{1}{12\pi}C\left(\frac{1}{C}\right)_{,\hat{u},\hat{u}},\label{hatuu}\\
 \langle T_{\hat{v}\hat{v}}\rangle 
  &=& -\frac{1}{12\pi}C\left(\frac{1}{C}\right)_{,\hat{v},\hat{v}}\label{hatvv}\\
 \langle T_{\hat{u}\hat{v}}\rangle &=& \frac{{\cal{R}}C^2}{96\pi},
 \label{hatuv}
\end{eqnarray}
where $\cal R$ is the two dimensional scalar curvature. 

We are now in the situation to
compute the exact expression for the quantum stress tensor     
in the two-dimensional self-similar collapse. 
As usual, we require that the regular center is given by 
$\hat{u}=\hat{v}$ and that
$\hat{v}$ coincides with the standard Eddington-Finkelstein
advanced time coordinate $v$.
Assuming the relation between internal and external null 
coordinates, 
\begin{equation}
 U=\alpha(u) \hspace{1cm} v=\beta(V)
\end{equation}
we can obtain all the coordinate relations as
\begin{equation}
 \hat{v}=v=\beta(V) \hspace{1cm} \hat{u}=\beta(U)
  =\beta(\alpha(u)). \label{relation}
\end{equation}
Then we can transform the quantum stress tensor given by
$(\hat{u},\hat{v})$ coordinates to the ones in the interior and the
exterior double null coordinates.

Using the coordinate relation equation (\ref{relation}) we can transform
the equations\ (\ref{hatuu}) - (\ref{hatuv}) to the components expressed in 
the interior $(U,V)$ coordinates and in the exterior $(u,v)$ coordinates.
The components of the interior $(U,V)$ coordinates is given by
\begin{eqnarray}
 \langle T_{UU}\rangle &=& F_{U}(\beta ')- F_{U}(A^2) \label{TUU}\\
 \langle T_{VV}\rangle &=& F_{V}(\beta ')- F_{V}(A^2) \label{TVV}\\
 \langle T_{UV}\rangle &=& -\frac{1}{24 \pi}(\ln A^2)_{,U,V} \nonumber\\
          &=& \pm \frac{1}{24\pi}
 \frac{z^2-R'^2}{2R'}\frac{d^2R'}{dz^2}  ~~~~~(z+R'\gtrless 0)
\label{TUV}
\end{eqnarray}
where 
\begin{equation}
 F_x(y) = \frac{1}{12\pi}\sqrt{y}\left(\frac{1}{\sqrt{y}}\right)_{,x,x}.
\end{equation}
The $\beta '$ in equation (\ref{TUU}) should be considered as
$\beta '(U)$.
After a long calculation, we obtain
\begin{eqnarray}
 \label{F_V}
 F_V(\beta') &=& \frac{1}{12\pi}\left\{\frac{3}{4}\left(\frac{(\beta')_{,V}}{\beta'}\right)^2 -\frac{1}{2}\frac{(\beta')_{,V,V}}{\beta'}\right\}  \nonumber\\
             &=& \frac{1}{12\pi}\frac{1}{4(1+x)^2} 
             \left\{(1+x-\frac{x^2}{2})^2
               +x^5(1+\frac{x}{2})\left(\frac{1}{\lambda}
            -\frac{2}{3x^3}+\frac{x}{\lambda}+\frac{1}{3x^2}\right)\right\}
\end{eqnarray}
where $x$ is determined by the equation
\begin{equation}
 \label{(r_b-v)/2M}
 \frac{r_b-v}{2M} = \frac{2}{3x^3}+\frac{2}{x}-\frac{1}{x^2}
                    -2\ln\frac{1+x}{x} .
\end{equation}
$F_U(\beta')$ can be expressed by the same equation (\ref{F_V})
but with $x$ which 
is determined by the equation related to the retarded time as
\begin{equation}
 \label{(r_b-beta)/2M}
 \frac{r_b-\beta(\alpha(u))}{2M} = 
        \frac{2}{3x^3}+\frac{2}{x}-\frac{1}{x^2}-2\ln\frac{1+x}{x} .
\end{equation}
$F_U(A^2)$ and $F_V(A^2)$ become
\begin{eqnarray}
 \label{F_U(A^2)}
 F_U(A^2) &=& \frac{1}{12\pi}\left\{\frac{1}{4}\left
 ( \frac{dR'}{dz}-1\right)^2 + \frac{1}{2}\frac{z^2 -
 R'^2}{2R'}\frac{d^2R'}{dz^2}\right\} \\
 \label{F_V(A^2)}
 F_V(A^2) &=& \frac{1}{12\pi}\left\{\frac{1}{4}\left( \frac{dR'}{dz}+1\right)^2 + \frac{1}{2}\frac{z^2 - R'^2}{2R'}\frac{d^2R'}{dz^2}\right\}. 
\end{eqnarray}

In the exterior Schwarzschild region we obtain 
\begin{eqnarray}
 \langle T_{uu}\rangle &=& - F_{u}(D^2)+\alpha'^2 F_{U}(\beta ')+ F_{u}(\alpha') \label{Tuu}\\
 \langle T_{vv}\rangle &=& - F_{v}(D^2) \label{Tvv}\\
 \langle T_{uv}\rangle &=& -\frac{1}{24 \pi}(\ln D^2)_{,u,v} \label{Tuv}\nonumber\\
          &=& \frac{1}{24\pi}\left(2\frac{M^2}{R^4}-\frac{M}{R^3}\right).
\end{eqnarray}
The $\beta '$ in equation (\ref{Tuu}) should be also considered as
$\beta '(U)$. 

In this region we obtain
\begin{equation}
 \label{F_u=F_v}
 F_u(D^2)=F_v(D^2)=-\frac{1}{24\pi}\left(\frac{3}{2}\frac{M^2}{R^4}
                    -\frac{M}{R^3}\right).
\end{equation}
These values are always positive for $R>\frac{3}{2}M$.
Also we obtain
\begin{eqnarray}
\label{F_u}
 F_u(\alpha') &=&
  \frac{1}{12\pi}\left[-\frac{\alpha'^2}{4}\left(1-\frac{w^2}{2\lambda}+\frac{w}{3}\right)^2+\frac{\alpha'}{4\left(2M\right)}\left(\frac{w^4}{3}+\frac{2w^8}{\lambda}-\frac{2w^7}{\lambda}-\frac{w^5}{3}\right)+\frac{1}{16\left(2M\right)^2}\left(8w^7-7w^8\right)\right] \nonumber \\
  &=&\frac{1}{12\pi}\frac{3w^6\lambda(-(5w^4+12w^3-8w^2-24w+12)\lambda
               -6w^7+18w^6-12w^5)}{16(2M)^2(3w^3-2\lambda-3w^4-w\lambda)^2}
\end{eqnarray}
where $w$ is determined by the equation
\begin{equation}
 \label{r_b-u/2M}
 \frac{r_b-u}{2M} = \frac{2}{3w^3}+\frac{2}{w}+\frac{1}{w^2}
                    +2\ln\frac{1-w}{w} .
\end{equation}
The relation between $u$ and $U$ is obtained as
\begin{equation}
 \frac{1}{\alpha'(u)}= - \frac{2M}{1-w}\left(\frac{1}{\lambda}
                       -\frac{2}{3w^3}-\frac{w}{\lambda}-\frac{1}{3w^2}\right).
\end{equation}

Here we interpret the outgoing flux emitted from the star.
For this sake, we look into the right hand side of equation~(\ref{Tuu}).
The first term clearly denotes the vacuum polarization,
which tends to vanish so rapidly as $R$ goes to infinity
that there is no net contribution to the flux at infinity.
The second term is originated in the interior of the star
as seen in equations (\ref{TUU}) and (\ref{TVV}).
First it appears as ingoing flux at the passage of the 
ingoing rays through the stellar surface, crosses the center and 
becomes outgoing flux.
Since this term depends on both $\alpha^{\prime}$
and $\beta^{\prime}$,
not only the outgoing but also ingoing null rays 
are relevant to this term.
It implies that this term strongly depends on the 
details of the spacetime geometry in the interior of the star.  
Since the third term depends only on $\alpha^{\prime}$,
only the outgoing null rays determine this contribution. 
Such a term is not seen in equation (\ref{TUU}).
Therefore, this contribution seems to originate at the passage of
the outgoing rays through the stellar surface.
As will be discussed later, the third term corresponds to the
Hawking radiation, while the second term becomes important
in the naked singularity explosion.

\section{Exact behavior of quantum stress tensor}
\label{sec:Exact behavior}
Using the above exact expressions, \footnote{
The work here is based on treatment such as those in ref \cite{Barve:1998tv}.
One of the referees pointed out some
problems with such an approach. In this footnote we quote his/her
remarks from the original report: (i) There are reasons to believe that
the approach is still not mathematically complete, for example, the
metric (\ref{UVmetric}) is degenerate on the CH. It is quite possible that the
divergence observed on CH is only a manifestation of this singularity
in the metric. In order for it to be taken seriously, one should work
with an interior metric which is non-singular on the CH, relating it
properly with the exterior coordinate system, which is not achieved
so far. (ii) The coordinate invariance of the results such as
divergence on CH is to be properly established. (iii) Approaches such
as above rely on Hawking like treatment applied to the naked singular
case. However, the situation here is significantly different when
spacetime contains a naked singularity. Then the null infinity may be
destroyed due to the presence of naked singularity. Then such a
calculation has to be abandoned, especially if there are radiations
escaping away from the strong gravity regions just prior to the epoch
of naked singularity formation. In fact, calculations such as those
of Ford and Parker \cite{Ford:1978ip}, Hiscock et al
\cite{Hiscock:1982pa} are careful on this point
and did consider only the marginal cases when the horizon and the
singularity coincided, and then cutting off the spacetime. In
references such as \cite{Barve:1998tv}, this important point seems to have been
ignored, and the results may be unreliable to that extent.}
we can now investigate the exact 
behaviors of the quantum stress tensor for the massless scalar 
fields.  The key processes are to get $x$ which satisfy equations\
(\ref{(r_b-v)/2M}) and (\ref{(r_b-beta)/2M}) and $w$ in equation
(\ref{r_b-u/2M}). 

From equations (\ref{uv}), (\ref{R*}) and (\ref{T}) it is easy to  confirm that
the following equations hold at the dust surface,
\begin{eqnarray}
 \frac{r-v}{2M} &=&
 \frac{2}{3}\left(\frac{R}{2M}\right)^{3/2}+2\sqrt{\frac{R}{2M}}-\frac{R}{2M}+2\ln\left(\frac{1}{\sqrt{\frac{R}{2M}}+1}\right) \\
\frac{r-u}{2M} &=&
 \frac{2}{3}\left(\frac{R}{2M}\right)^{3/2}+2\sqrt{\frac{R}{2M}}+\frac{R}{2M}-2\ln\left(\frac{1}{\sqrt{\frac{R}{2M}}-1}\right).
\end{eqnarray}
Therefore $x$ in equations(\ref{(r_b-v)/2M}) and (\ref{(r_b-beta)/2M}) and
$w$ in equation (\ref{r_b-u/2M}) become 
\begin{equation}
 x, w = \sqrt{\frac{2M}{R}}
\end{equation}
where $R$ is the value at the dust surface on the corresponding null
trajectory.  Therefore,
to obtain the value of $x$ in equation (\ref{(r_b-v)/2M}) 
at some spacetime point, 
we integrate the ingoing null geodesic equation in the past direction
from there to the surface of the cloud. 
For equation (\ref{(r_b-beta)/2M}) at first we 
integrate the outgoing null geodesic equation in the past direction
to the center. From there we integrate the ingoing null geodesic
equation in the past direction to the surface of the cloud.
These integration is performed numerically by the 4-th order
Runge-Kutta method.
We need $w$ in the exterior of the dust cloud. 
At the surface $u$ is obtained from equations (\ref{uv}),
(\ref{R*}) and (\ref{T}). Therefore we should not integrate the null
geodesic equation in the exterior.

To determine what would be actually measured, the world line 
of the observer must be specified. 
For the observer with the velocity $u^\mu$, the energy density 
$\langle T_{\mu\nu}\rangle u^\mu u^\nu$ and energy 
flux $\langle T_{\mu\nu}\rangle u^\mu n^\nu$ are 
measured, where $u^\mu n_\mu=0$. 

\begin{figure}
  \begin{center}
    \leavevmode
    \begin{tabular}{c}
    \subfigure[]{\epsfysize=240pt\epsfbox{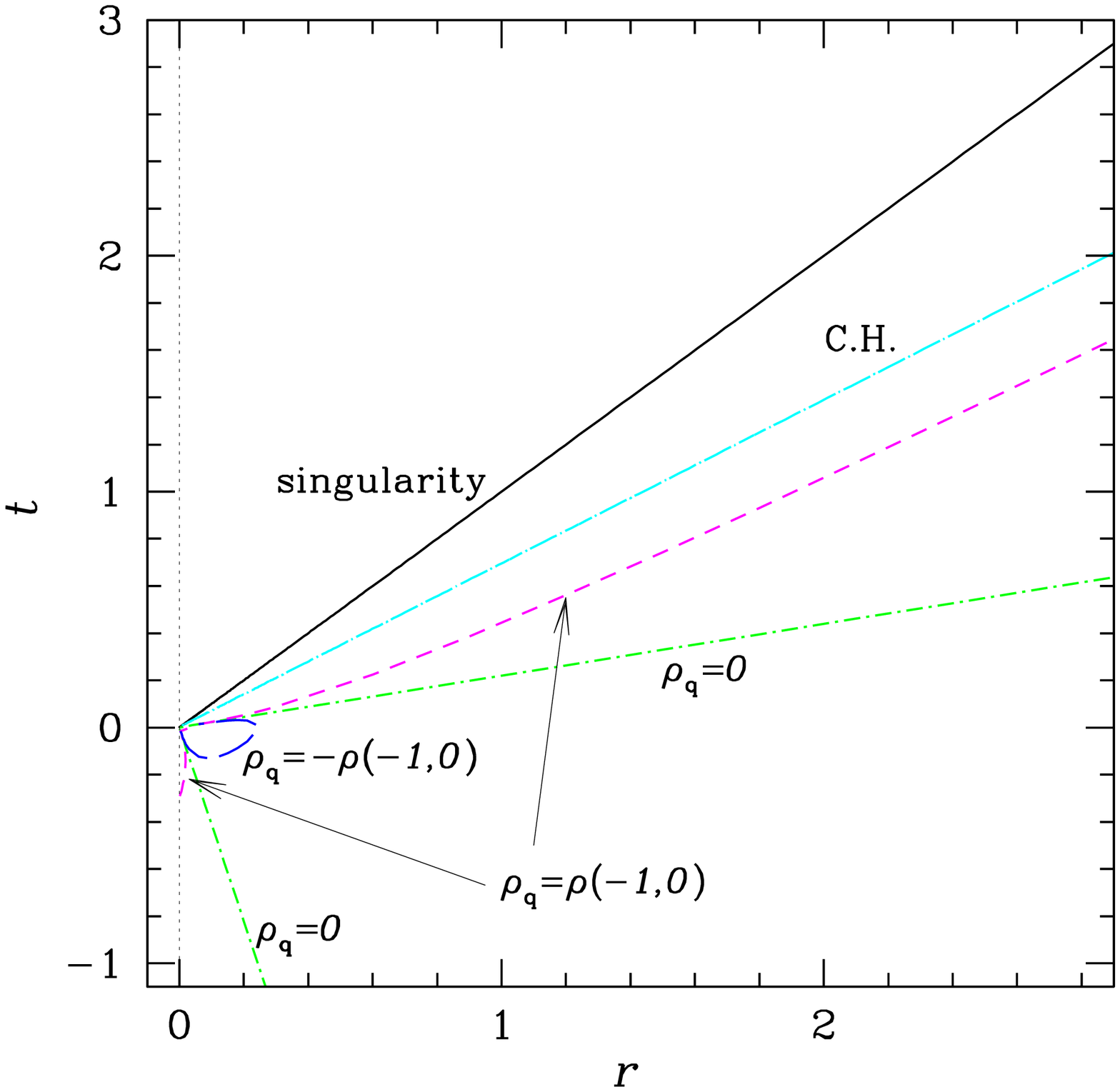}}  
    \subfigure[]{\epsfysize=240pt\epsfbox{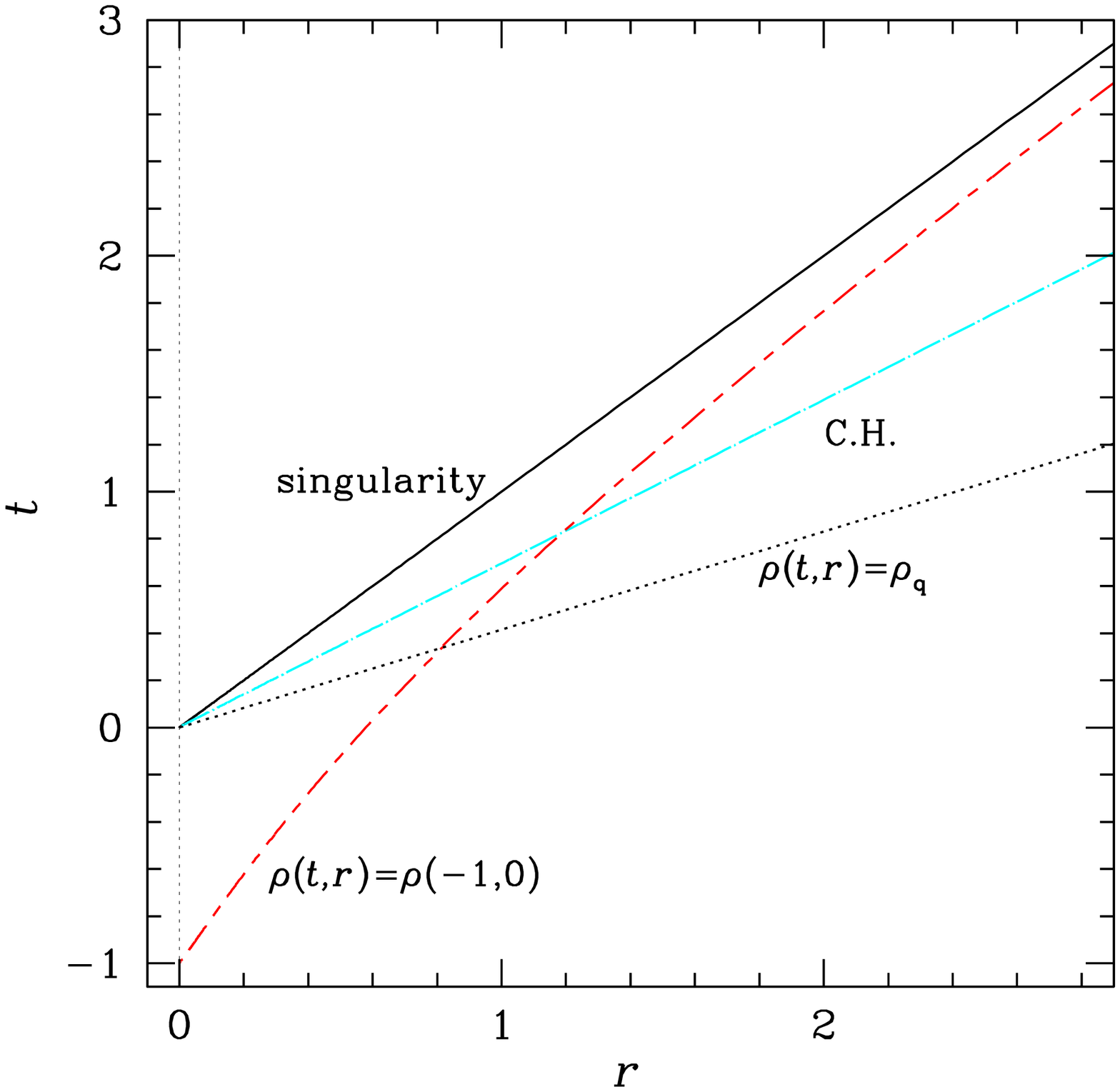}}  
    \end{tabular}  
  \caption{Plots of the lines $\rho_{\mbox{\scriptsize q}}=$ const in
  (a) and $\rho(t,r)=$ const and $\rho_{\mbox{\scriptsize
  q}}=\rho(t,r)$ in (b). Each line has the following meaning.
  Solid line: singularity, long-dashed-dotted line: Cauchy horizon,
  dashed line: $\rho_{\mbox{\scriptsize q}}=\rho(-1,0)$, dashed-dotted line:
  $\rho_{\mbox{\scriptsize q}}=0$, long-dashed line:
  $\rho_{\mbox{\scriptsize q}}=-\rho(-1,0)$, long-dashed-short-dashed line:
  $\rho(t,r)=\rho(-1,0)$, dotted line: $\rho_{\mbox{\scriptsize q}}=\rho(t,r)$.
    }
 \label{fig:density}
  \end{center}
 \end{figure}

To investigate the importance of the back reaction for the central
singularity formation, we compare the energy density observed by the
comoving observer to the background energy density around the center.
The energy density observed by the comoving observer becomes
\begin{equation}
 \label{rho_{q}}
 \rho_{\mbox{\scriptsize {q}}}\equiv\langle T_{\mu\nu}\rangle u^\mu u^\nu 
         = \frac{\langle T_{UU}\rangle }{r^2(z-R')^2}
                        +\frac{\langle T_{VV}\rangle }{r^2(z+R')^2}
                        \mp 2 \frac{\langle T_{UV}\rangle}{r^2(z^2-R'^2)}.
\end{equation}
The results are shown in figure  
\ref{fig:density}. Basically we use the background parameters
$\lambda=0.1$ and $r_b=10^3$. The lines of
$\rho_{\mbox{\scriptsize {q}}}=$ const are plotted in (a) and the line of
$\rho(t,r)=$ const is in (b).  These two values coincide with each other on
the dotted line in (b). Below this line the background density is
larger than the quantum energy density. 
We see especially that near the center the background density is larger than 
the  energy density of the scalar field until the central
singularity. Therefore we conclude that  the
back reaction does not become significant during the semiclassical
evolution. 

We can show that $\rho_{\mbox{\scriptsize q}}$ does not overcome
$\rho(t,r)$ at the center for naked case $\lambda \le 0.115427\cdots$.
Substituting equations\ (\ref{R}), (\ref{R'}), (\ref{dR'/dz}) and
(\ref{d^2R'/dz^2}) into equations\ (\ref{F_U(A^2)}), (\ref{F_V(A^2)}) 
and (\ref{TUV}) and taking the limit $r\rightarrow 0$, we obtain 
\begin{equation}
 \label{F_U(A^2)=F_V(A^2)}
 F_U(A^2)=F_V(A^2)=\frac{7}{432\pi}
\end{equation}
and
\begin{equation}
 \langle T_{UV}\rangle=\mp \frac{1}{216\pi}.
\end{equation}
Since at the center $v=\beta(\alpha(u))$ is holds, 
\begin{equation}
 \label{F_U(beta')=F_V(beta')}
 F_U(\beta')=F_V(\beta').
\end{equation}
The denominators of right hand side of equation (\ref{rho_{q}}) become
\begin{equation}
 \label{t^2}
 r^2(z-R')^2=r^2(z+R')^2=r^2(z^2-R'^2)=t^2
\end{equation}
at the center. Using these equations we obtain
$\rho_{\mbox{\scriptsize q}}$ at 
the center as
\begin{equation}
 \rho_{\mbox{\scriptsize q}}=\frac{1}{t^2}\left(2F_V(\beta')-\frac{5}{216\pi}\right).
\end{equation}
The range of the $F_V(\beta')$ at the center is estimated in Appendix
\ref{app:estimate}.
We conclude that $\rho_{\mbox{\scriptsize q}}$ is less than $\rho(t,0)$ at
the center as
\begin{equation}
 \rho_{\mbox{\scriptsize q}}(t,0) \leq \frac{1}{54\pi t^2}<\frac{1}{6\pi t^2}=\rho(t,0).
\end{equation}
One can also show that $\rho_{\mbox{\scriptsize q}}$ is not negative
at the center.

\begin{figure}
  \begin{center}
    \leavevmode
    \epsfysize=250pt\epsfbox{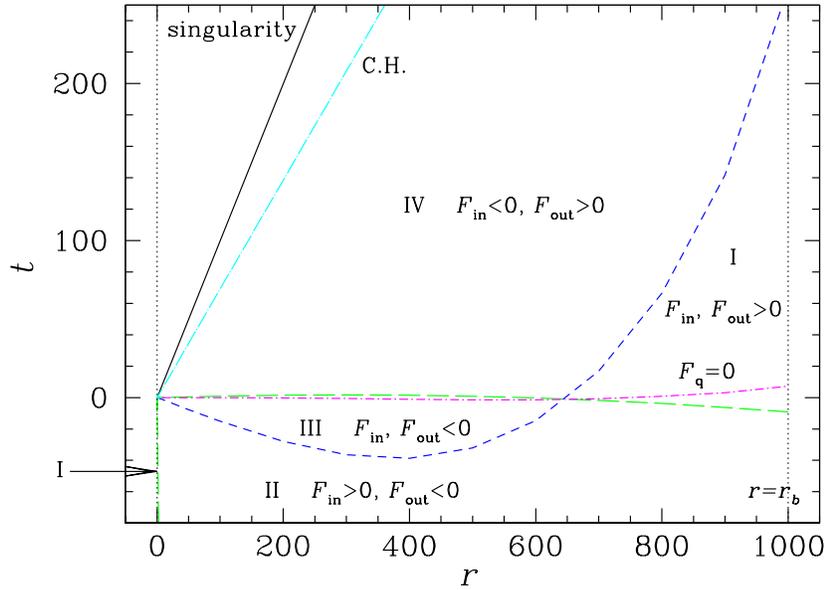}    
  \caption{
   The solid line and long-dashed-dotted line correspond to the singularity 
  and the Cauchy horizon, respectively. The dashed line denotes the
  line where the ingoing part of flux vanishes. On the long dashed lines
  the outgoing part vanishes. Region I, II, III, and IV
  correspond to positive in and out flux, positive in and
  negative out flux, negative in and out flux, and negative in
  and positive out flux, respectively. On the short-dashed-dotted line 
  the observed net flux vanishes.
   }
 \label{fig:region}
  \end{center}
 \end{figure}

Next we consider the energy flux measured by the comoving observer.
For this observer, energy flux becomes
\begin{equation}
 \label{eq:flux}
 F_{\mbox{\scriptsize q}}\equiv \langle T_{\mu\nu} \rangle u^\mu n^\nu 
 = \frac{\langle T_{UU}\rangle}{r^2(z-R')^2}
                        -\frac{\langle T_{VV} \rangle}{r^2(z+R')^2}.
\end{equation} 
In the interior region, outgoing flux is proportional to 
$\langle T_{UU}\rangle$
and ingoing flux is proportional to $\langle T_{VV}\rangle$. 
Here we define the ingoing part $F_{\mbox{\scriptsize in}}$ and the outgoing
part $F_{\mbox{\scriptsize out}}$ of the flux as
\begin{equation}
 F_{\mbox{\scriptsize in}}\equiv \frac{\langle
 T_{VV}\rangle}{r^2(z+R')^2}, ~~~~~
 F_{\mbox{\scriptsize out}}\equiv
 \frac{\langle T_{UU}\rangle}{r^2(z-R')^2}.
\end{equation}
Following the sign of these two values we divide the 
interior region into four parts. Region I: positive in and out
flux, region II: negative out flux and positive in flux, region III:
negative in and out flux, and  region IV: positive out flux and negative in 
flux. The results are shown in figure \ref{fig:region}.
 Flux observed by a timelike observer becomes zero in region I
and III. The observed flux vanishes at $r=0$ and dotted dashed
line in figure \ref{fig:region}. 
We can show that the flux becomes zero at the center. From equations 
(\ref{F_U(A^2)=F_V(A^2)}) and (\ref{F_U(beta')=F_V(beta')}) we obtain
\begin{equation}
 \langle T_{UU} \rangle = \langle T_{VV} \rangle.
\end{equation}
Using this equation and equation (\ref{t^2}) we see that the right hand
side of equation (\ref{eq:flux}) is equal to zero at the center. Also
we see 
\begin{equation}
 \langle T_{UU} \rangle = \langle T_{VV} \rangle > \frac{1}{60\pi}-\frac{7}{432\pi} > 0.
\end{equation}
Therefore in and out part of the flux are positive at the regular center.

\begin{figure}
  \begin{center}
    \leavevmode
    \epsfysize=280pt\epsfbox{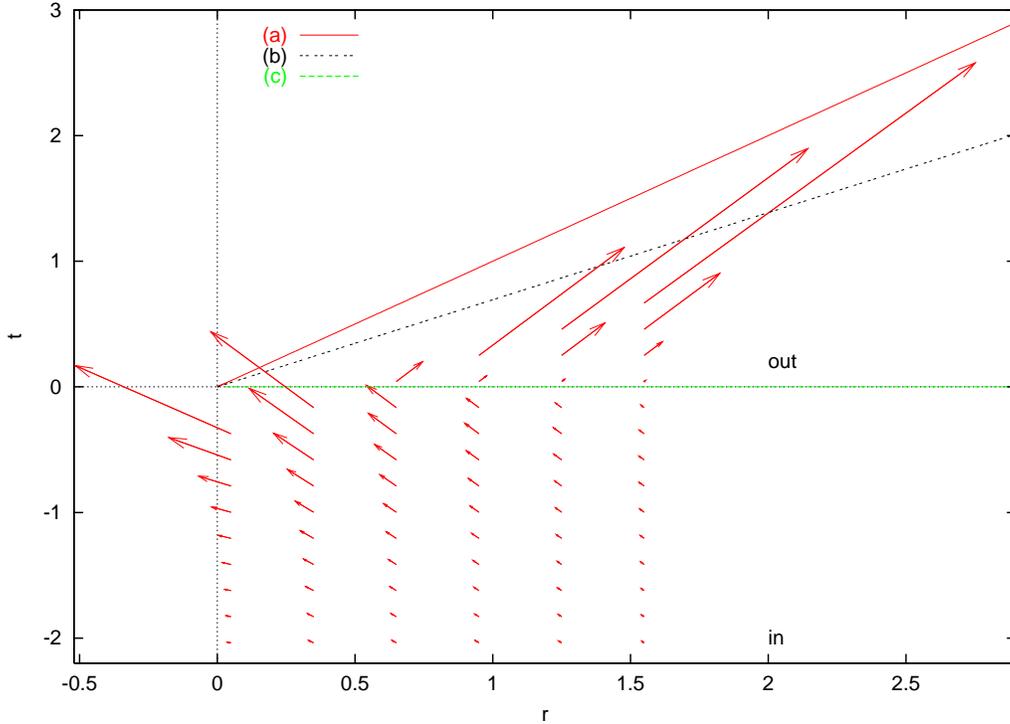}    
  \caption{
   Schematic figure for the flux observed by comoving observer.
  Lines (a) and (b) correspond to the singularity and the Cauchy
  horizon, respectively. Line (c) denotes the line where the observed
  flux vanishes.
   }
 \label{fig:flux}
  \end{center}
 \end{figure}

We show this flux schematically in figure \ref{fig:flux}. The intensity 
is proportional to the length of arrows along the $r$-axis. The comoving 
observers receive inward flux first and then outward flux after the
crossing of the line (c). The
intensity grows as the Cauchy horizon is approached. 

\begin{figure}
  \begin{center}
    \leavevmode
    \epsfysize=280pt\epsfbox{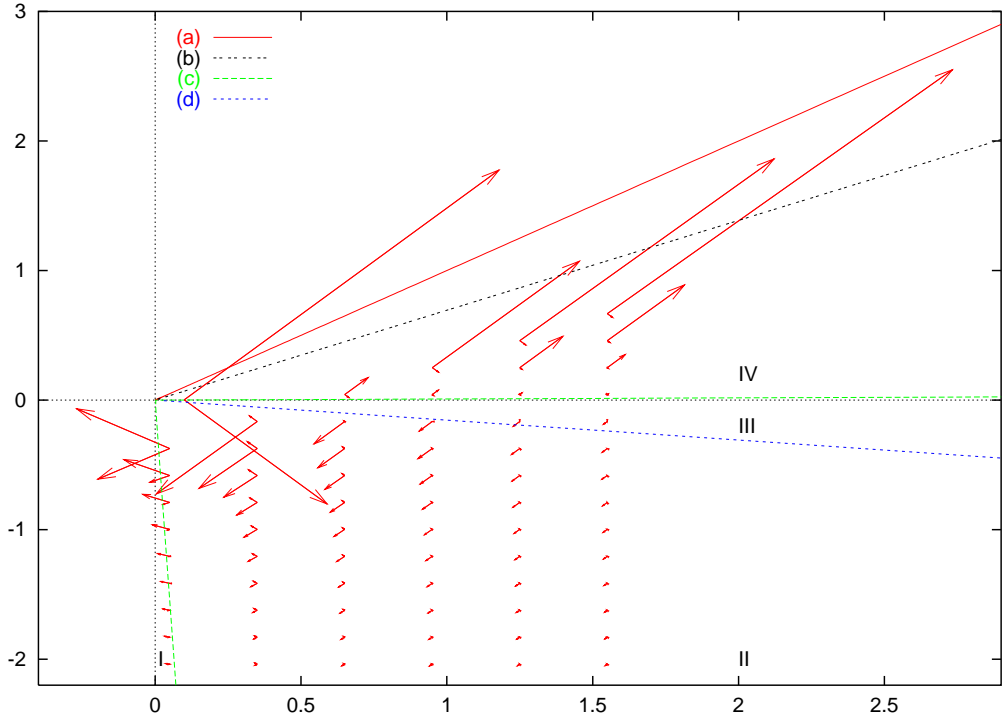}    
  \caption{
   Schematic figure for the in and outgoing parts of flux observed 
  by comoving observer.  Lines (a) and (b) correspond to the
  singularity and the Cauchy horizon, respectively, Lines (c) and (d)
  denote the lines where  outgoing part and ingoing part vanish 
  respectively. 
  Region I, II, III, and IV are similar to the ones in figure \ref{fig:region}.
   }
 \label{fig:nflux}
  \end{center}
 \end{figure}

In figure \ref{fig:nflux}, we show in and outgoing parts of the flux
schematically. The intensity 
is proportional to the length of arrow along the $r$-axis.
The future and past directed arrows correspond to the positive and
negative flux respectively. 
The outgoing part has large intensity near the Cauchy
horizon. The ingoing part has large intensity only near the central
singularity. 

\begin{figure}
  \begin{center}
    \leavevmode
    \begin{tabular}{c}
    \subfigure[]{\epsfysize=240pt\epsfbox{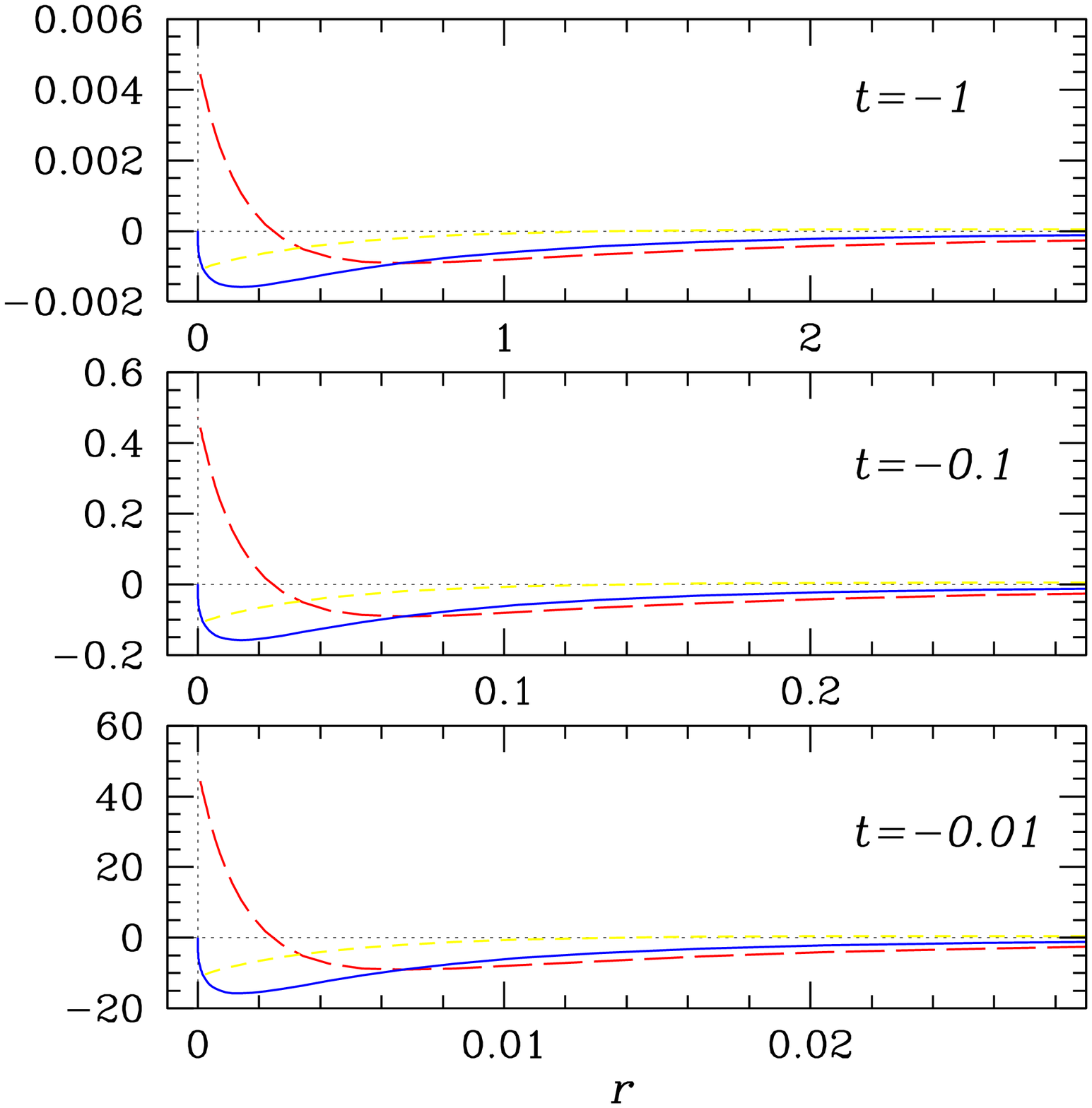}}  
    \subfigure[]{\epsfysize=240pt\epsfbox{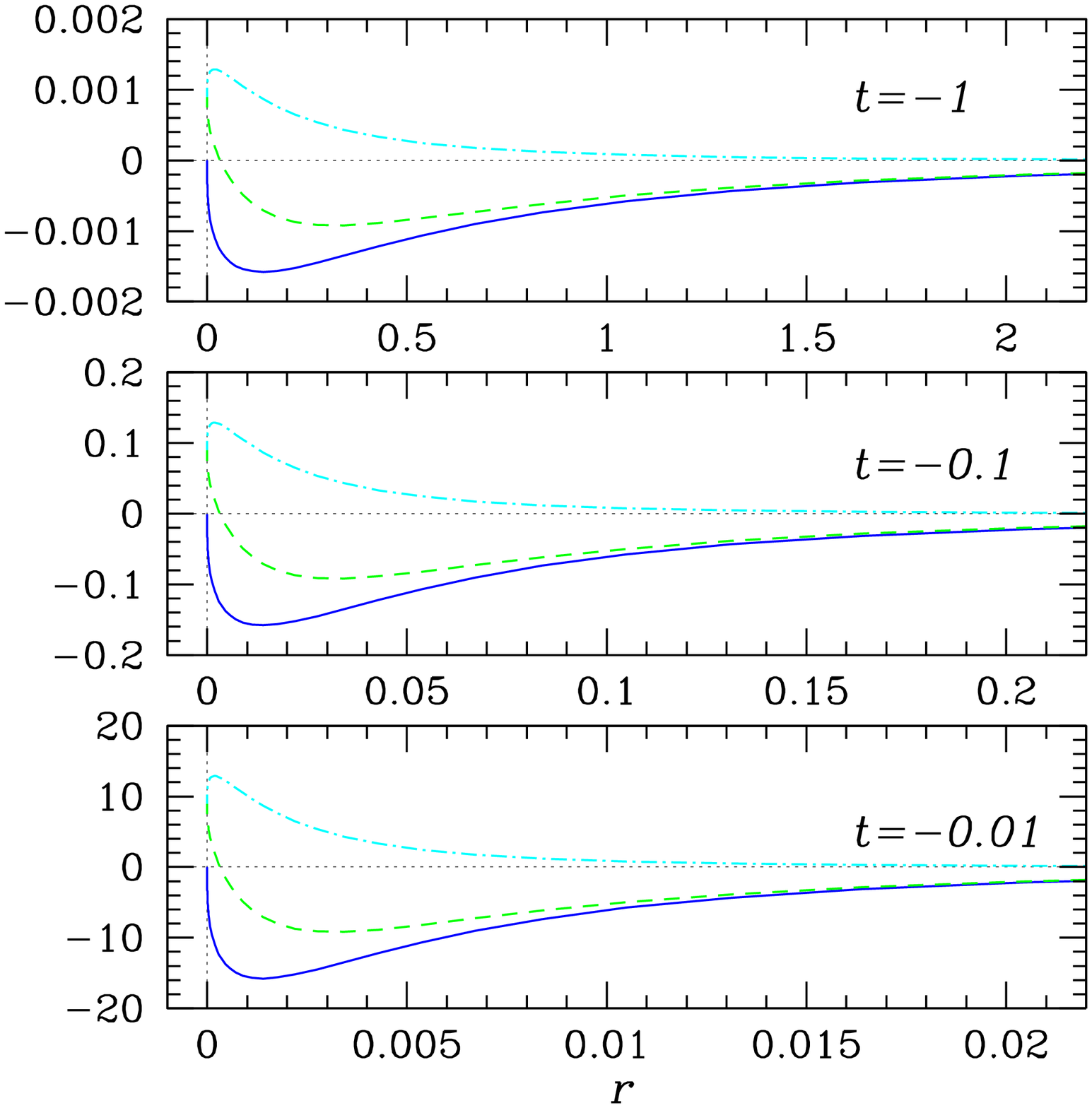}}  
    \end{tabular}  
  \caption{
    Plots of the quantum variables at $t=-1, -0.1, -0.01$.    
   In (a)  long-dashed lines denote $\rho_{\mbox{\scriptsize q}}$, solid
   lines denote $F_{\mbox{\scriptsize q}}$, and dashed lines denote
  quantum pressure $\langle T_{\mu\nu}\rangle n^\mu n^\nu$.
   In (b) solid lines denote flux $F_{\mbox{\scriptsize q}}$, dashed 
  lines denote outgoing parts of flux $F_{\mbox{\scriptsize out}}$, 
  and dashed-dotted lines
  denote ingoing parts of flux $F_{\mbox{\scriptsize in}}$.
   }
 \label{fig:t<0}
  \end{center}
 \end{figure}

\begin{figure}
  \begin{center}
    \leavevmode
    \begin{tabular}{c}
  \subfigure[$t=0$]{\epsfysize=200pt\epsfbox{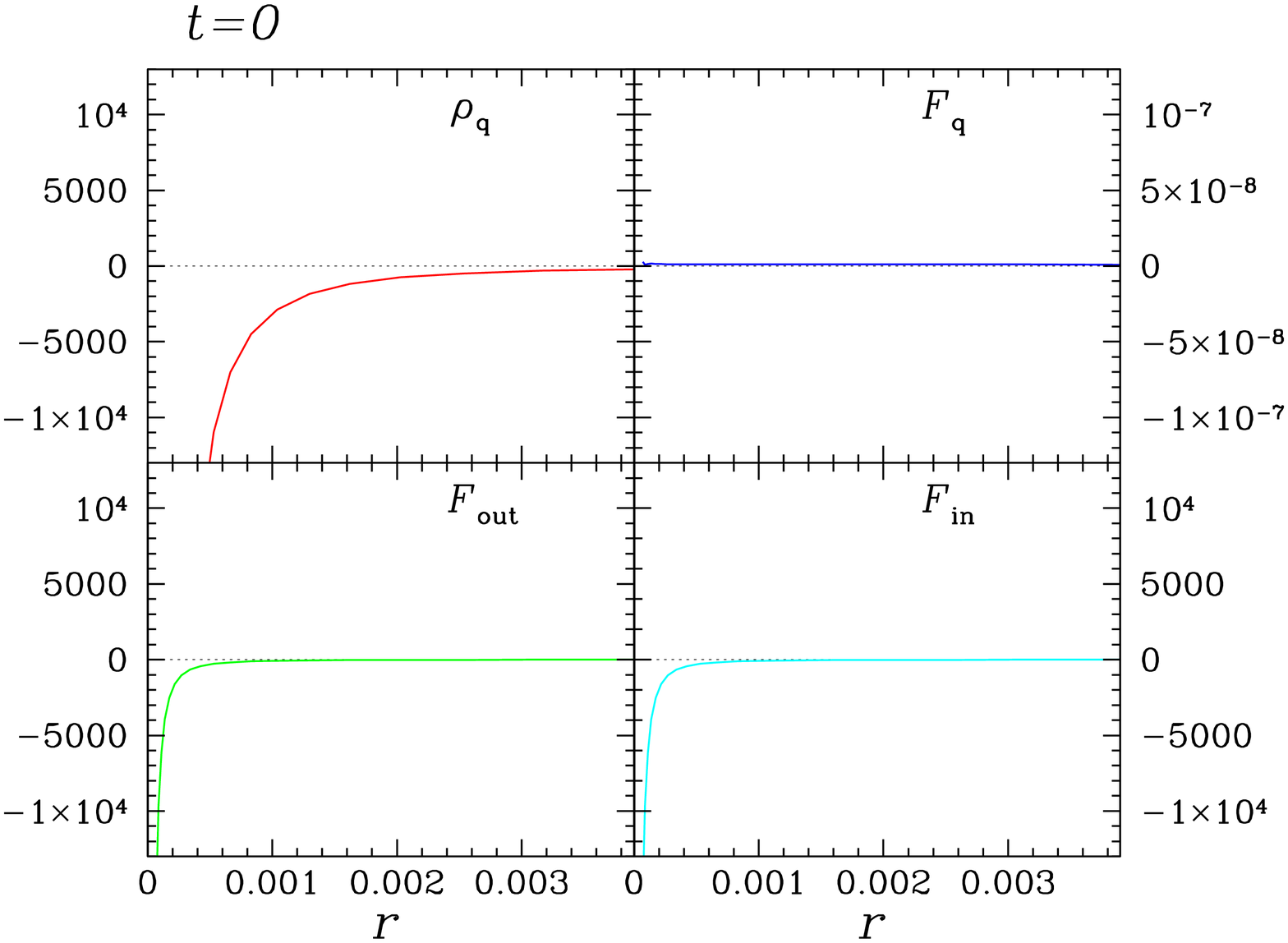}}  
  \subfigure[$t=0.01$]{\epsfysize=200pt\epsfbox{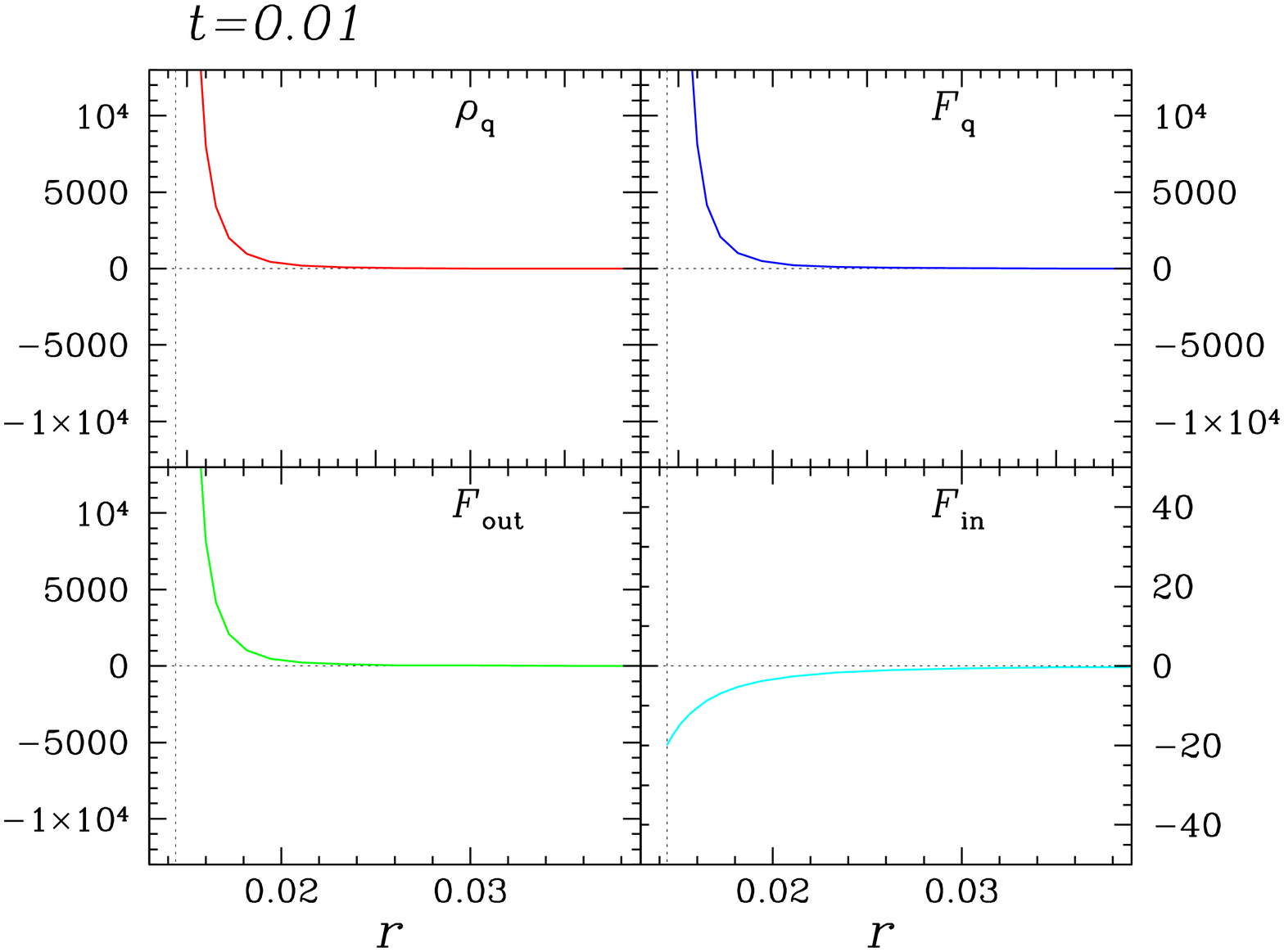}}  \\
  \subfigure[$t=0.1$]{\epsfysize=200pt\epsfbox{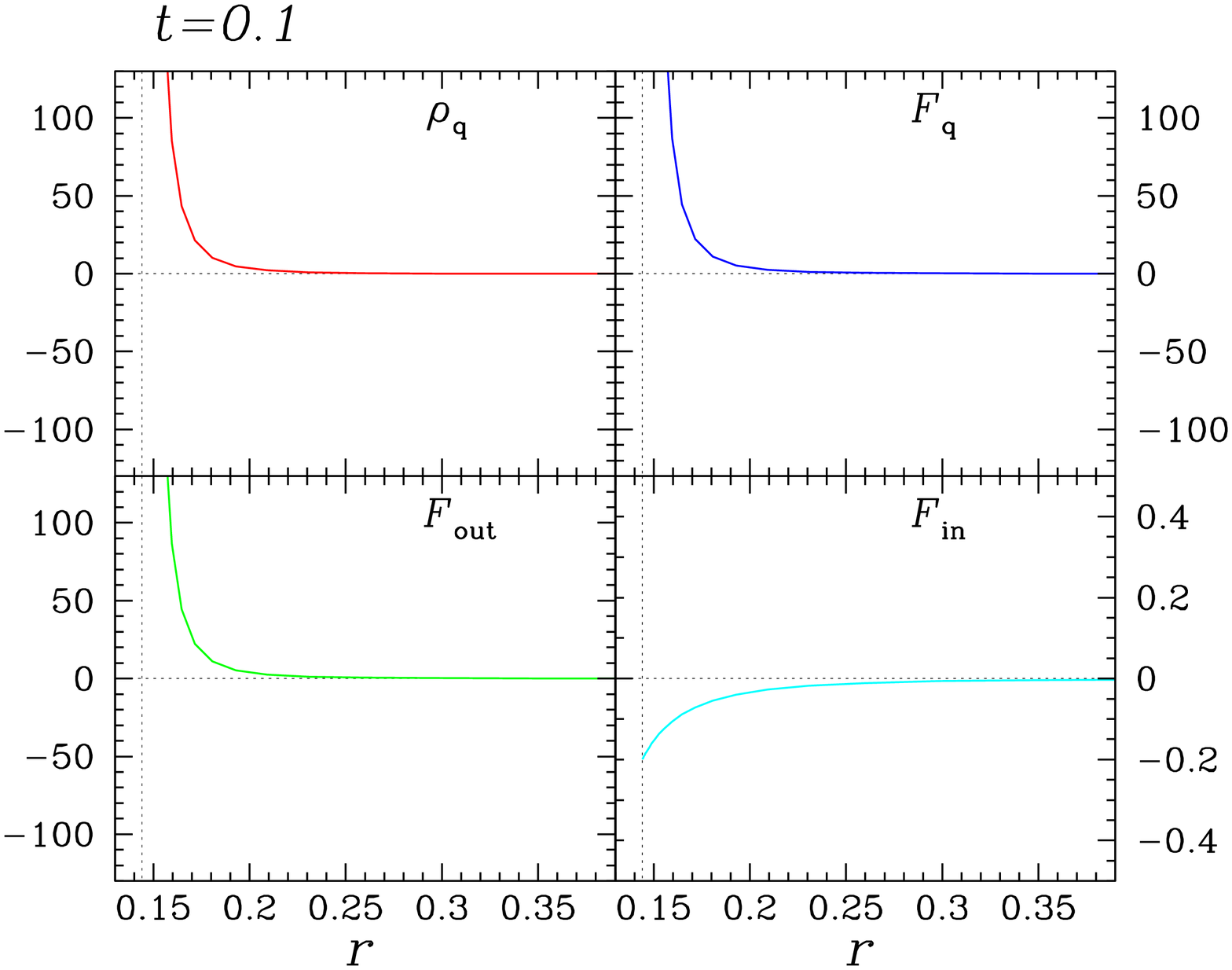}}  
  \subfigure[$t=1$]{\epsfysize=200pt\epsfbox{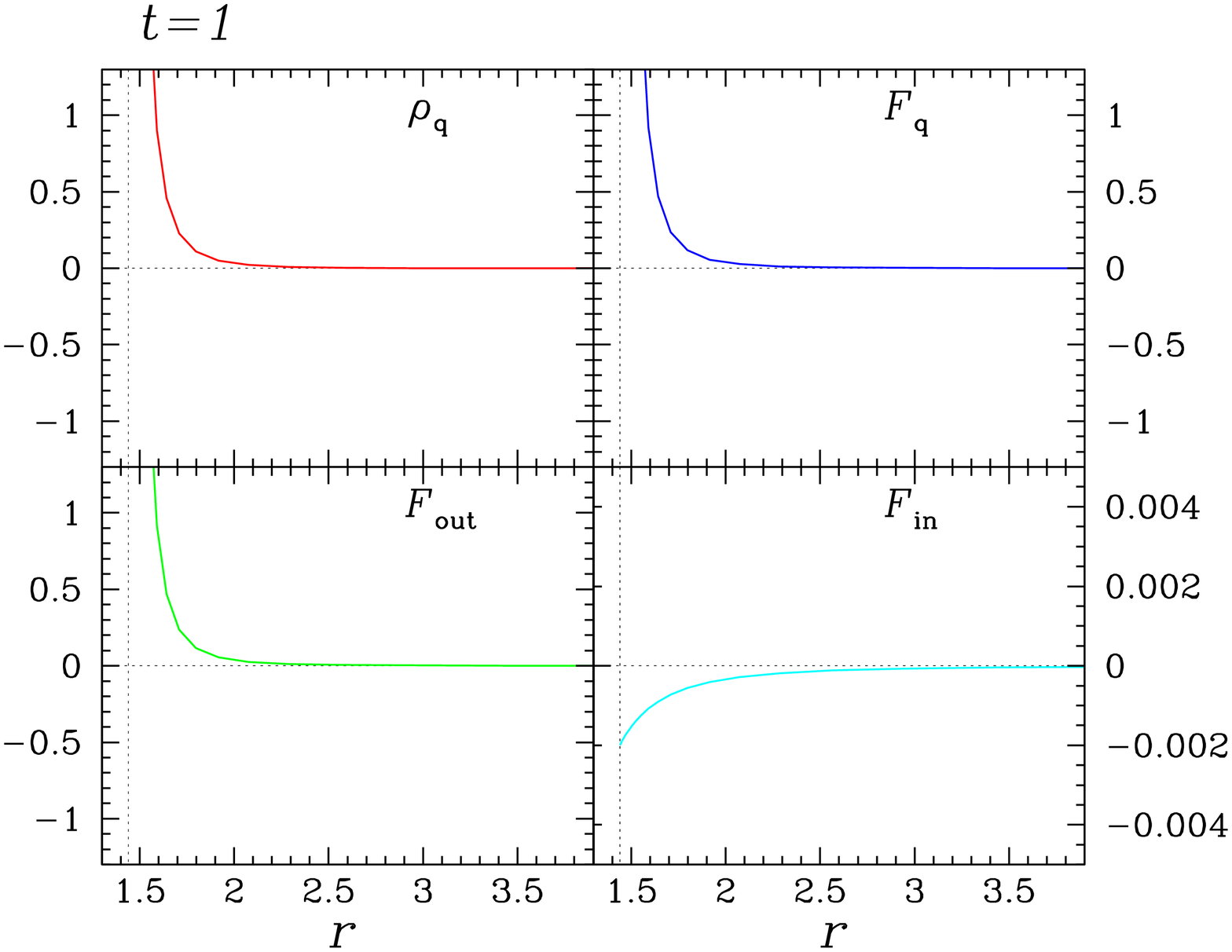}}  
    \end{tabular}  
  \caption{
 Plots of the quantum variables at $t=0, 0.01, 0.1, 1$ in (a), (b),
  (c), and (d) respectively. In each subfigure upper-left panel shows
  $\rho_{\mbox{\scriptsize q}}$, upper-right panel shows
  $F_{\mbox{\scriptsize q}}$, lower-left panel shows
  $F_{\mbox{\scriptsize out}}$, and lower right panel shows
  $F_{\mbox{\scriptsize in}}$. Vertical dotted lines in (b), (c) and
  (d) represent the Cauchy horizon.
   }
 \label{fig:t>0}
  \end{center}
 \end{figure}

We plot these quantum values at $t=$ const for $t<0$ in figure \ref{fig:t<0}
and for $t \ge 0$ in figure \ref{fig:t>0}. Figure \ref{fig:t<0} (a)
shows quantum energy density, energy flux, and pressure $\langle
T_{\mu\nu}\rangle n^\mu n^\nu$ measured by
comoving observer.  The energy density grows positively near the
center surrounded by negative energy density. The
$F_{\mbox{\scriptsize q}}$ is negative,
i.e., the inward energy flow and vanishes at the center.
In figure \ref{fig:t<0} (b), in and outgoing parts of flux are
plotted. This inward quantum energy flow is constructed mainly by 
positive ingoing
and negative outgoing flux. At the center the positive outgoing flux 
cancel out the positive ingoing flux, where the net flux is zero. 

In figure \ref{fig:t>0} (a) we plot the quantum values at $t=0$. The energy
density negatively diverges as the center is approached. The in- and
outgoing flux also negatively diverge as the center is
approached. These two almost cancel each other.
Figures \ref{fig:t>0} (b), (c) and (d) denote the behaviors of quantum
values for $t=0.001, 0.1, 1$. In this region the energy density 
and the flux diverge positively as the Cauchy horizon is
approached. The divergency of the flux is originated from the
divergence of the positive outgoing flux. The ingoing flux is negative
and finite.

\begin{figure}
  \begin{center}
    \leavevmode
    \epsfysize=300pt\epsfbox{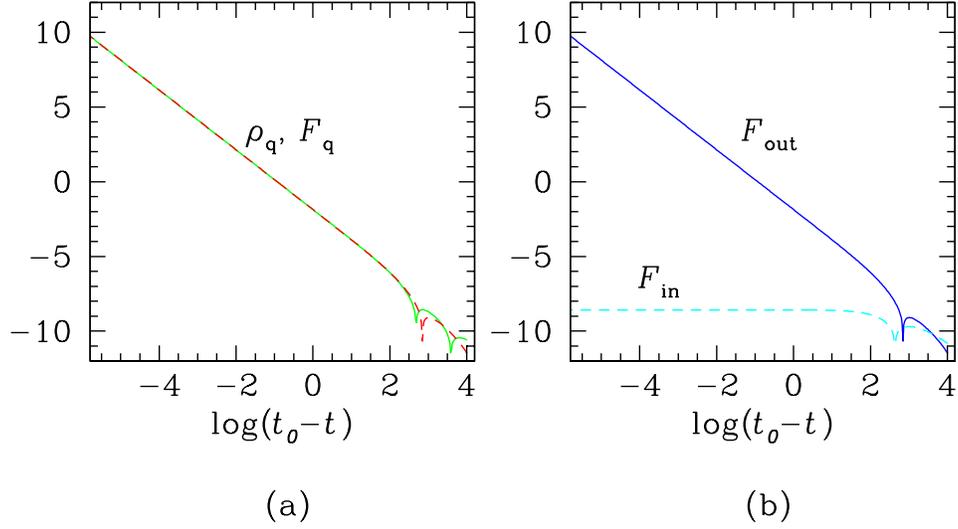}    
  \caption{
   Logarithmic plots of the absolute values of quantum variables at the
  surface $r=r_b$. 
   }
 \label{fig:surface}
  \end{center}
 \end{figure}

The same values at the surface $r=r_b$ are plotted in figure \ref{fig:surface}.
In figure \ref{fig:surface} (a) energy density and flux are plotted. In (b) 
in and out parts are plotted. There are inverse square divergence of
the energy density, flux and out going part of the flux.

\begin{figure}
  \begin{center}
    \leavevmode
    \epsfysize=300pt\epsfbox{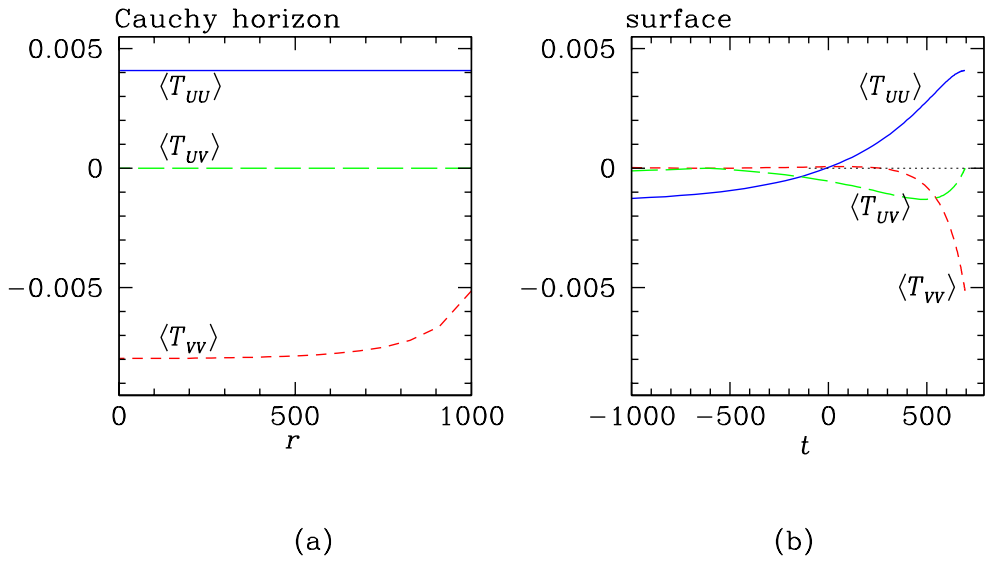}    
  \caption{
 Plots of $\langle T_{UU}\rangle$, $\langle T_{VV}\rangle$, and
  $\langle T_{UV}\rangle$, at (a) Cauchy horizon, and at (b) $r=r_b$. 
   }
 \label{fig:sandc}
  \end{center}
 \end{figure}

We plot $\langle T_{UU} \rangle$, $\langle T_{VV}\rangle$, and
$\langle T_{UV}\rangle$ along the Cauchy
horizon in figure \ref{fig:sandc} (a) and along $r=r_b$ in (b). 
We can confirm that there are positive outgoing flux along the Cauchy
horizon and negative ingoing flux across it.  

Here we summarize the behavior of quantum field in the interior of
the star. At first inward flux appears and then the positive energy 
is gathered near the center surrounded by slightly negative energy region. 
As the collapse proceeds the central positive energy grows and
concentrates smaller region. 
At the naked singularity formation the
gathered positive energy is converted to the outgoing positive diverging
flux.  The left negative energy goes down across the
Cauchy horizon.  It will be reach the spacelike singularity.

\begin{figure}
  \begin{center}
    \leavevmode
    \epsfysize=250pt\epsfbox{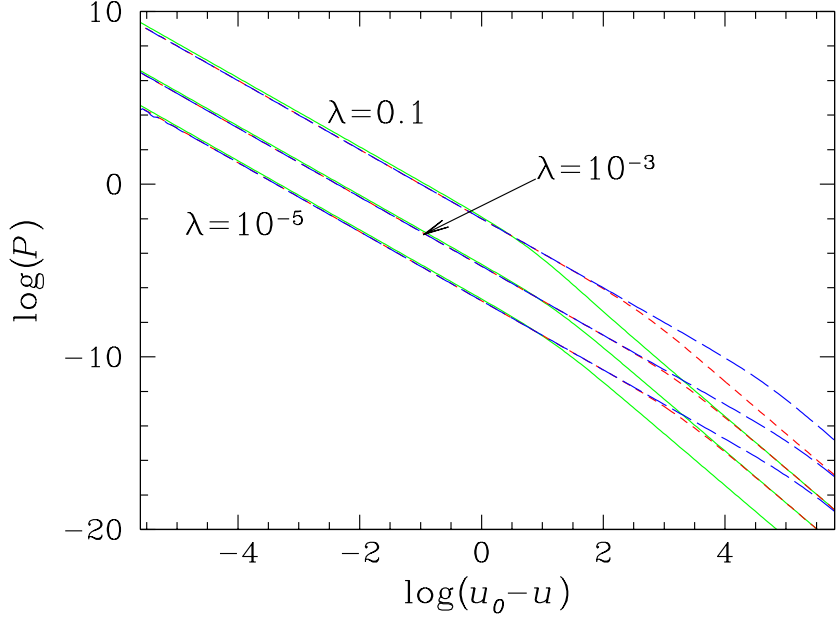}    
  \caption{
   Radiated power received at the infinity. Upper lines correspond to
  $\lambda=0.1,$ , and middle lines correspond to $10^{-3}$, and lower 
  lines correspond to $10^{-5}$. 
  Solid, dashed, and long-dashed lines
  correspond to $r_b=10, 10^3, 10^5$, respectively.
   }
 \label{fig:infty_flux}
  \end{center}
 \end{figure}

Next we consider the total radiated  energy received at infinity.
The radiated power of quantum flux can be expressed as
\begin{equation}
 \label{P}
 P = \langle T_{\mu\nu} \rangle u^\mu n^\nu = 
    \alpha'^2 F_U(\beta') + F_u(\alpha')
\end{equation}
where $u^\mu$ is considered as the velocity of static observer.
We plot this in figure \ref{fig:infty_flux}. We reproduce the results
of the geometric optics estimates for the 4D model, i.e., 
the inverse square
dependence for the retarded time in the approach to the Cauchy horizon.
In this stage the behavior of $P$ is determined by $\lambda$ and 
weakly depends on $r_b$. On the other hand, during the early times,  
the $P$ is proportional to the total mass $M=\frac{1}{2}\lambda r_b$
and grows as $(u_0-u)^{-3}$. This behavior agrees with the minimally
coupling case for $C^{\infty}$ background \cite{Harada:2000ar}.

\begin{figure}
  \begin{center}
    \leavevmode
    \epsfysize=250pt\epsfbox{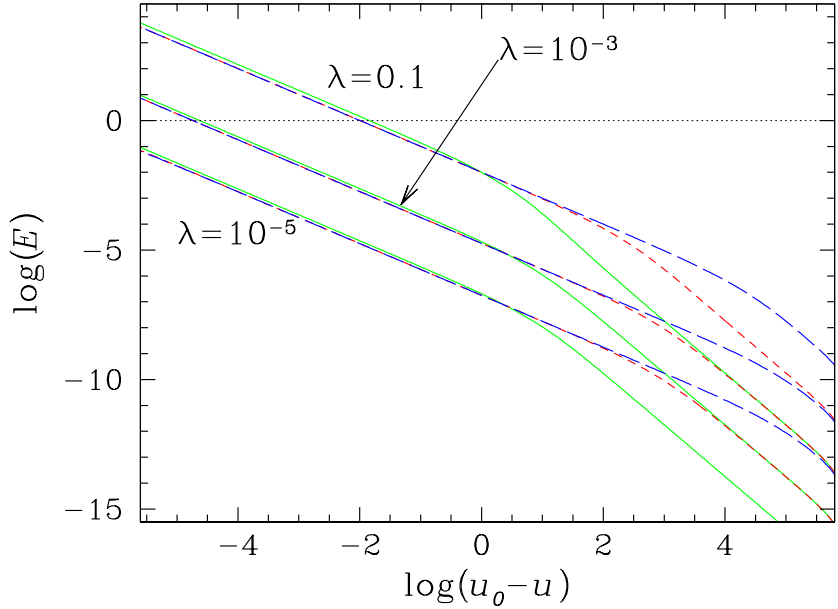}    
  \caption{
   Integrated radiated power. Upper lines correspond to
  $\lambda=0.1,$ , and middle lines correspond to $10^{-3}$, and lower 
  lines correspond to $10^{-5}$. 
  Solid, dashed, 
  and long-dashed lines correspond to $r_b=10, 10^3, 10^5$, respectively.
   }
 \label{fig:tflux}
  \end{center}
 \end{figure}

The total energy is obtained from the integration of $P$ with respect to 
the retarded time, 
\begin{equation}
  E  = \int (\alpha'^2 F_U(\beta') + F_u(\alpha')) du
\end{equation}
at the future null infinity. The results are shown in figure
\ref{fig:tflux}. We see that the total emitted energy is
much smaller than the mass of the original star in the range that the
semiclassical approximation can be trusted. 

\begin{figure}
  \begin{center}
    \leavevmode
    \epsfysize=240pt\epsfbox{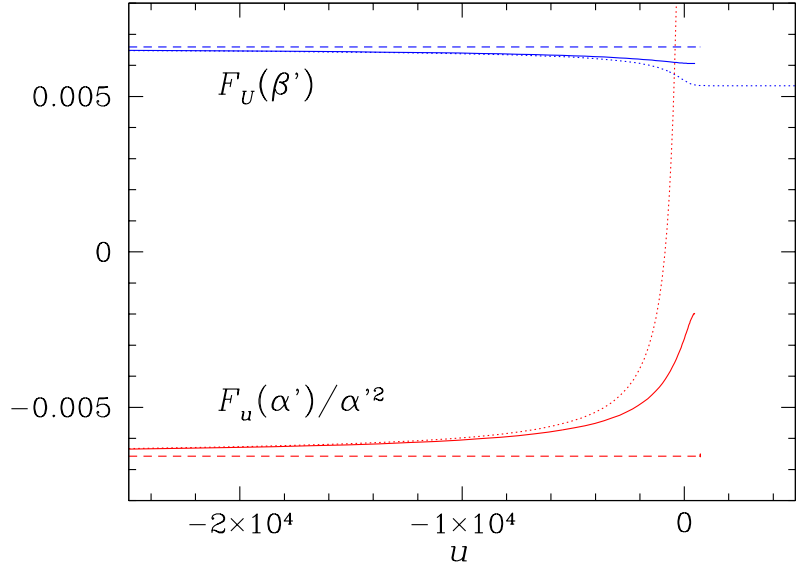}    
  \caption{
   Plots of $F_U(\beta')$ for upper lines, and
  $F_u(\alpha')/\alpha'^2$ for lower lines. Solid lines correspond to 
  $\lambda=0.1$ and $r_b=10^3$ which is plotted to Cauchy horizon
  $u\approx 479.0390$, and dashed lines correspond to
  $\lambda=10^{-3}$ and $r_b=10^5$, which is plotted to the Cauchy
  horizon $u\approx 715.3728$. Dotted lines denote a black hole
  case $\lambda=1$ and $r_b=10^2$. In this case the retarded time at
  event horizon is $u=\infty$.
   }
 \label{fig:comp}
  \end{center}
 \end{figure}

If we compare the $\alpha'^2F_U(\beta')$ and $F_u(\alpha')$
we can say something about the origin of the diverging flux.
In figure \ref{fig:comp} we plot $F_U(\beta')$ and
$F_u(\alpha')/\alpha'^2$. It is seen that the first term of
equation (\ref{P}) is positive and the second term is negative. As was
shown in \cite{Barve:1998tv} $\alpha'^2$ behaves like 
\begin{equation}
 [z_{\mbox{\scriptsize out}}(w)-z_{\mbox{\scriptsize out}}(w \mbox{ at the Cauchy horizon})]^{-2}
\end{equation}
where 
\begin{equation}
 z_{\mbox{\scriptsize out}}(w)=1-\frac{2\lambda}{3w^3}.
\end{equation}
The first term of equation (\ref{F_u})  negatively diverges in the 
approach to the Cauchy horizon.
Therefore we conclude that the positive divergence of the radiated 
power is originated from the term $\alpha'^2 F_U(\beta')$ which has
strong dependence on the spacetime geometry in the interior of the star. 

Let us turn to the inspection for the black hole
formation. $F_U(\beta')$ and $F_u(\alpha')/\alpha'^2$ for this case
are also plotted in figure \ref{fig:comp}. For the larger $u$ there
appear the difference between the naked singularity case and the covered
one. This difference mainly depend on the behavior of $\alpha'$. In the
approach to the event horizon, $w$ goes to unity and then $\alpha'$
goes to zero. It is easy to see that only the second term of 
the radiated power is left in this limit and it becomes
\begin{equation}
F_u(\alpha')=\frac{1}{192\pi\left(2M\right)^2}.
\end{equation}
This means that $F_u(\alpha')$ is interpreted as the Hawking radiation 
contribution. From equation (\ref{F_u=F_v}) we obtain
\begin{equation}
 \langle T_{vv}\rangle = -\frac{1}{192\pi\left(2M\right)^2}
\end{equation}
at the event horizon. This ingoing negative flux crossing into the
black hole balances with the energy loss by the positive Hawking flux.

There is similar balance between in- and outgoing flux at the Cauchy
horizon for the naked singularity explosion. In the asymptotic region 
positive diverging flux is observed in the approach to the Cauchy
horizon. It has been shown above that there is negative ingoing flux
crossing the Cauchy horizon. This negative flux diverge only at the
central naked singularity. This diverging negative ingoing flux is 
balanced with the diverging positive outgoing flux which propagates 
along the Cauchy horizon.

At the last of this section we give a speculation for some possible
scenario of the practical quantum field radiation at the final stage
of the collapse inspired by the above analysis.
Practically the classical naked singularity should be 
replaced by an observable high-density region. 
After the appearance of this high-density
region, the surrounding dust fluid falls in this 
region as positive energy flow. On the other hand quantum negative
flux flows in this region. The balanced positive outgoing flux
emerges from the cloud and propagates towards infinity.
If the intensity of the classical positive energy flow is always equal
to the one of the quantum 
negative ingoing flux, then the central high-density region would keep the
steady state until all envelope disappears. 
It may be argued that in this case the inflow of the dust into the
central region is converted to the outgoing quantum energy flux.
Therefore we would observe the emission with finite
duration rather than the instantaneous explosion. 
And finally the small high-density core would be left at the 
center.  
Even after this modification, it can be expected that the
observational aspects of this radiation is still different from the
Hawking radiation.

\section{Summary}
\label{sec:Summary}

We have investigated the expectation value of quantum stress tensor 
for the massless scalar field on the 2D self-similar LTB spacetime.
If we apply the semiclassical treatment up to the Cauchy horizon,
we can depict the behavior of the energy flux inside the star in case of 
the naked singularity explosion as follows. 
As the dust collapse proceeds the quantum field flows inward and the
positive energy is accumulated in the center surrounded by slightly negative 
energy. At the naked singularity formation this gathered positive
energy is converted to the diverging outgoing flux. The negative energy
envelope flows inward crossing the Cauchy horizon. The diverging
outgoing flux emerges from the stellar surface and propagates along
the Cauchy horizon. 
There is energy balance relation between ingoing and outgoing flux at the
Cauchy horizon similar to the Hawking radiation. However this balance
is not long-standing like the Hawking radiation but instantaneous. 
The ingoing negative 
flux diverges at the central naked singularity, which is balanced with
the outgoing diverging flux along the Cauchy horizon. 

In the above picture of diverging flux, however,  
we omit a limitation of semiclassical treatment. 
The high density region very near the naked singularity and 
the part of spacetime which is causally connected to this high density 
region would not be exactly described only by the classical gravity.
We need a quantum gravity theory there. Therefore the semiclassical 
treatment is not plausible there.    
In the suitable region for the 
semiclassical treatment the energy density of the quantum field is
less than the background energy density near the center. 
Therefore we can conclude that the back reaction of the quantum
radiation is insignificant to the naked singularity formation.
It can be said that quantum gravitational effects are more
important.

In the practical point of view the naked singularity would be replaced 
by the sufficiently high density region (e.g., Planck density). In this 
situation the naked singularity explosion would be also replaced by a
practical event. 
If we perform the replacement of the naked singularity, 
the inflow of the dust into the central region would be converted to
the outgoing quantum energy flux in the assumption of the equilibrium 
of the classical and quantum inflows.
As a result, we could expect that
the outgoing flux would be milder and have longer time
duration than the naked singularity explosion, in which the flux 
is emitted divergingly in less than a Planck time.

\section*{Acknowledgment}
We are grateful to T P Singh and H Kodama for helpful discussions and
comments. HI partially performed this work at the Osaka University.
He is grateful to M Sasaki, F Takahara and the other members of
the theoretical astrophysics group of Osaka University for their
useful comments and continuous encouragement.
TH is grateful to K Maeda for his continuous encouragement.
This work was partially supported by the Grant-in-Aid for Scientific
Research (No. 05540) from the Japanese Ministry of Education, Science, 
Sports, and Culture.

\appendix
\section{Estimate of $F_V(\beta')$}
\label{app:estimate}
To estimate the range of $F_V(\beta')$ at the center we rewrite it as
\begin{equation}
 F_V(\beta') = \frac{1}{48\pi}\left\{\left(1-\frac{x^2}
                                {2\left(1+x\right)}\right)^2
             +\frac{x^2\left(1+\frac{x}{2}\right)\left(x-2\right)}
                  {3\left(1+x\right)^2}
             +\frac{x^5\left(1+\frac{x}{2}\right)}{\lambda\left(1+x\right)}
             \right\}. \label{A1}
\end{equation}
As we have quoted above, $x$ is represented as
\begin{equation}
 x=\left.\sqrt{\frac{2M}{R}}\right|_{r=r_b,t=t_j}=\left(\frac{2}{3}
                    \frac{\lambda}{\left(1-z_j\right)}\right)^{1/3}
\end{equation}
where $t_j$ and $z_j$ means the values of the junction point. 
For the central time $t<0$, the related $z_j$ is always negative.
Therefore following relation holds,
\begin{equation}
 \frac{x^3}{\lambda}=\left(\frac{2}{3}\right)
                    \frac{1}{\left(1-z_j\right)}<\frac{2}{3}. \label{x^3/lambda}
\end{equation}
Using the upper bound for the $\lambda$ we obtain 
\begin{equation}
 0 \leq x < 0.426.
\end{equation}

First we consider upper bound of the $F_V(\beta')$. From equation
(\ref{x^3/lambda}) it can be shown that the bracket of equation (\ref{A1}) 
is less than
\begin{equation}
  f_1(x) \equiv \left(1-\frac{x^2}
                                {2\left(1+x\right)}\right)^2
             +\frac{x^2\left(1+\frac{x}{2}\right)\left(x-2\right)}
                  {3\left(1+x\right)^2}
             +\frac{2x^2\left(1+\frac{x}{2}\right)}{3\left(1+x\right)}
\end{equation}
for $0 \leq x < 0.426$. In this region $f_1(x)$ has a maximum value at 
$x=0$ as 
\begin{equation}
 f_1(x) \leq f_1(0)=1.
\end{equation}
As a results $F_V(\beta ')$ is bounded as
\begin{equation}
 F_V(\beta')\leq \frac{1}{48\pi}.
\end{equation}

Next we consider the lower bound of $F_V(\beta ')$. The third term of
the right hand side of equation (\ref{A1}) is not negative for $0 \leq x
< 0.426$. We define
\begin{equation}
  f_2(x) \equiv \left(1-\frac{x^2}
                                {2\left(1+x\right)}\right)^2
             +\frac{x^2\left(1+\frac{x}{2}\right)\left(x-2\right)}
                  {3\left(1+x\right)^2}.
\end{equation}
It can be easily shown that $f_2(x)$ is monotonically 
decreasing function for $0 \leq x < 0.426$. Therefore we obtain an
inequality 
\begin{equation}
  \frac{4}{5}< f_2(0.426)=0.819\cdots < f_2(x).
\end{equation}
Therefore $F_V(\beta')$ is bounded as
\begin{equation}
 \frac{1}{60\pi} < F_V(\beta') \leq \frac{1}{48\pi}
\end{equation}
for the naked case.

\end{document}